\documentclass[aps,prl,twocolumn,groupadress]{revtex4}

\usepackage{epsfig}
\usepackage{amsmath}

\begin{document}

\title{Coupling between quasiparticles and a bosonic mode in the normal state of HgBa$_2$CuO$_{4+\delta}$}
\author{Y. Gallais$^{1}$, M. Le Tacon$^{2,3}$, A. Sacuto$^{2,3}$ and D. Colson$^{4}$}
\address{$^1$ Department of Physics, Columbia University, New York, New York 10027, USA\\
  	$^2$ Laboratoire Mat\'eriaux et Ph\'enom\`enes Quantiques (UMR 7162 CNRS),
	Universit\'e Paris 7, 2 place Jussieu 75251 Paris, France\\
  	$^3$ Laboratoire de Physique du Solide, ESPCI, 10 rue
	Vauquelin 75231 Paris, France\\
  	$^4$ Service de Physique de l'Etat Condens\'{e}, CEA-Saclay,
	91191 Gif-sur-Yvette, France
}

\begin{abstract}
We report a doping dependent study of the quasiparticles dynamics in HgBa$_2$CuO$_{4+\delta}$ via Electronic Raman Scattering. A well-defined energy scale is found in the normal state dynamics of the quasiparticles over a broad doping range. It is interpreted as evidence for coupling between the quasiparticles and a collective bosonic mode whose energy scale depend only weakly with doping. We contrast this behavior with that of the superconducting gap whose amplitude near the node continuously decreases towards the underdoped regime. We discuss the implications of our findings on the nature of the collective mode and argue that electron-phonon coupling is the most natural explanation.
\end{abstract}

\maketitle

While a consensus has now been reached for a $d$-wave superconducting order parameter over a wide doping range 
for hole doped cuprates \cite{tsuei-revue}, the mechanism of high-$T_c$ superconductivity is not yet established. Originally, identification of the order parameter symmetry as $d$-wave was believed to settle down the issue in favor of a purely electronic pairing interaction because Coulomb repulsion between electron leads more naturally to $d$-wave pairing than conventional electron-phonon coupling. In recent years however, the electron-phonon scenario has regained attention partly due to Angle Resolved Photoemission Spectroscopy (ARPES) results showing evidence for strong coupling between the quasiparticles (qps) and a collective bosonic mode in  hole-doped cuprates \cite{Lanzara-nature,Kaminski,Bogdanov,Dresde}. Identification of the collective mode remains highly controversial since both phonon and spin fluctuations have been proposed to explain the observed anomalies \cite{shen-nagaosa,norman-chubukov}. The main candidate for the spin fluctuation scenario is the q=($\pi$,$\pi$) spin resonance observed in the superconducting state of most hole-doped cuprates \cite{bourges,Dai}. For the phonon scenario, both the out of phase O buckling and the Cu-O bond-stretching phonons have been considered \cite{dev-shen}. The identification of the leading coupling mechanism in the normal state of the cuprates is of crucial importance and could help reveal the microscopic mechanism of high-temperature superconductivity. While some optical conductivity data have been interpreted along these lines \cite{schachinger}, our current understanding relies heavily on the interpretation of ARPES measurements where an experimental concensus has yet to be reached. The core of the controversy concerns the existence, or not, of a well-defined energy scale in the normal state's quasiparticle dispersion. Indeed contrary to the phonon case, the contribution of the magnetic resonance mode is expected to be at best marginal above the transition temperature $T_c$.
 
In this letter we propose the point of view of Electronic Raman Scattering (ERS) on the controversy. ERS is a two-particle charge probe that allows the study of qps dynamics along both the anti-nodal ($(0,0)$-$(0,\pi)$) and the nodal ($(0,0)$-$(\pi,\pi)$) directions of the cuprate's Brillouin zone. Here we focus on the behavior of the 
qps dynamics along the nodal directions where, contrary to the anti-nodal directions, a coherent response can be observed even in the underdoped regime \cite{chen,Nemetschek,Gallais2005}. We investigate ERS data on single layer HgBa$_2$CuO$_{4+\delta}$ (Hg-1201) as a function of doping. We show the existence of an energy scale in the qps dynamics in the normal state over a wide doping range. We interpret this energy scale as evidence for the coupling between the qps and a collective bosonic mode and propose a simple model of ERS to account for our experimental findings. We discuss the possible candidates for this mode and find phonon to be the most plausible candidate.

\begin{figure*}
\centering
\epsfig{file=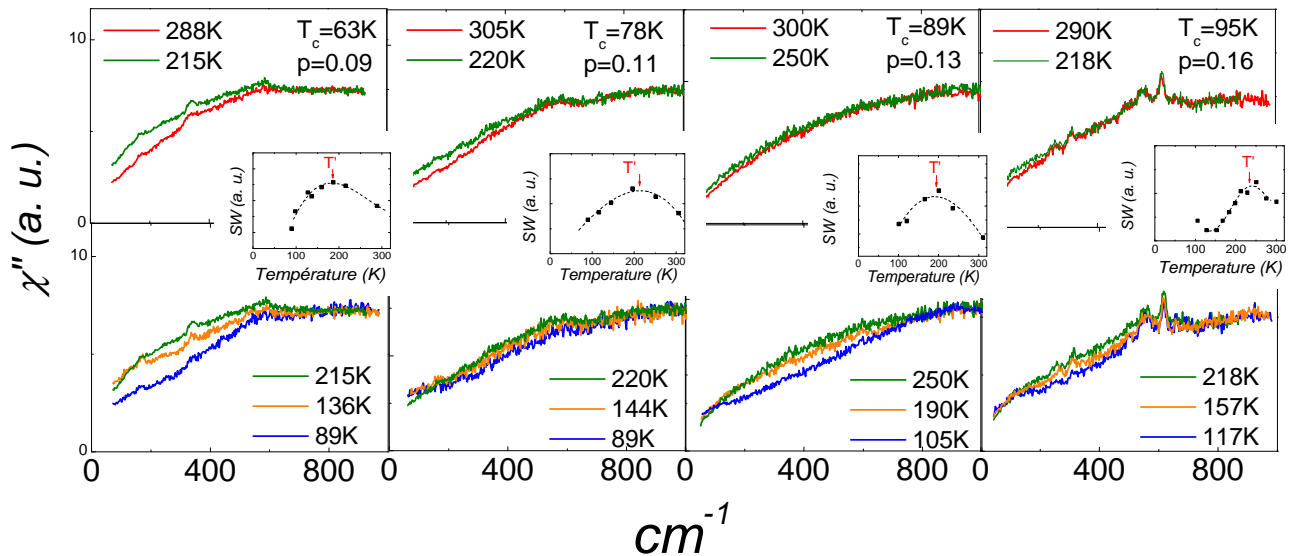, width=17cm,silent=}
\caption{Temperature dependence of the B$_{2g}$ ERS response $\chi^{"}$ as a function of doping in the normal state of Hg-1201. The upper panel shows the high temperature behavior ($T>T^{'}$) while the lower shows the low temperature one ($T<T^{'}$). $T^{'}$ is defined through the evolution of the integrated response (SW=$\int^{1000}_{70}\chi^{"}(\omega)d\omega$) shown in insets for each sample.}
\label{spectres}
\end{figure*}

The Hg-1201 single crystals studied here have been successfully grown
by the flux method \cite{Colson,Colson2}. The ERS measurements have
been performed on four as-grown single crystals with different
dopings. Their magnetically measured transition temperatures, $T_c$,
are 95~K, 89~K, 78~K and 63~K (transition widths are less than 5~K for all samples).
 We will refer to these
crystals as Hg95K, Hg89K, Hg78K and Hg63K respectively.
Their doping $p$ is deduced from the universal relationship between $\frac{T_c}{T_{c,max}}$ and $p$ \cite{Presland}. The first sample
is very close to optimal doping while the latter three are
underdoped. The spectra were obtained using the 514.5~nm
(2.4~eV) excitation line of an Ar$^+$-Kr$^+$ laser. The scattered
light was analysed using a triple grating spectrometer (JY-T64000)
equipped with a nitrogen cooled CCD detector. In this study we
focus on the B$_{2g}$ symmetry which can be selected by using crossed ingoing and outgoing polarisations along the Cu-O bonds of the CuO$_2$ plane (xy geometry). In the cuprates, the
B$_{2g}$ symmetry probes the electronic excitations along the nodal
directions. The spectra presented here have been corrected for the
spectral response of the spectrometer and divided by the Bose-Einstein
factor. They are thus proportional to the imaginary part of the
Raman response function $\chi^"$. All the referred temperatures
have been corrected for the estimated laser heating.

\begin{figure}
\centering
\epsfig{figure=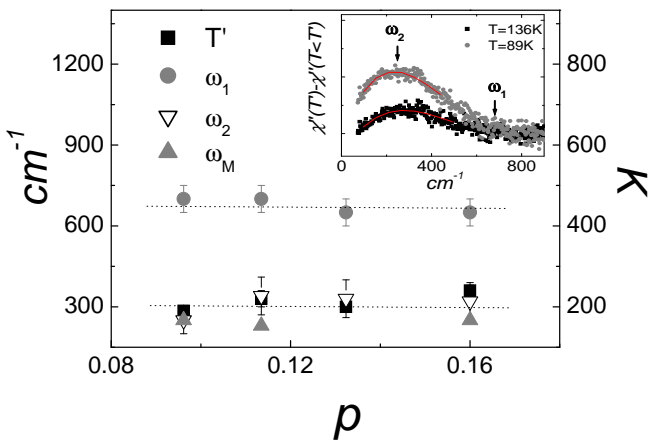, width=0.99\linewidth, clip=}
\caption{Summary of the evolutions of the temperature ($T^{'}$) and energy scales ($\omega_1$ and $\omega_2$) of the depletion as a function of doping $p$. The dotted lines are guides to the eye. The inset shows the difference spectra $\chi^{"}(T^{'})$-$\chi^"(T<T^{'})$ for two temperatures below $T^{'}$ for Hg63K. $\omega_1$ is defined as the onset of the depletion while $\omega_2$ is defined as its maximum. $\omega_M$ is the extracted characteristic frequency from our analysis (see further in the text).}
\label{evolution}
\end{figure}

In figure \ref{spectres}, we show the normal state Raman responses in the B$_{2g}$ symmetry as a function of temperature for the four samples between 70 and 1000 cm$^{-1}$. At high temperatures ($T>$200~K) and for all samples, the low frequency ($\omega<$500~cm$^{-1}$) response shows an increase of spectral weight (upper panel). The slope of the Raman response at vanishing frequencies being directly proportional to the qps static lifetime, the low frequency increase of spectral is a direct consequence of a metallic-like behavior (qps lifetimes increase as temperature is decreasing). Below approximately 200~K however, the response develops an anomalous loss of spectral weight with decreasing temperature (lower panel). Remarkably the depletion is seen for all four samples and starts in the same frequency range, around 600-700 cm$^{-1}$. The depletion however, does not extend to the lowest frequencies since the responses at lower temperature (see for example $T$=89~K and 140~K for Hg78K) cross again the higher ones ($T$=220~K) around 100 cm$^{-1}$, consistently with a metallic-like behavior at very low frequencies. Again this behavior is observed in all samples except for the most underdoped sample (Hg63K) where,the depletion is more pronounced and the responses most likely cross at a frequency below 70~cm$^{-1}$ (see the responses at 89 and 136~K). The appearance of the depletion can be tracked by plotting the temperature dependence of the integrated spectral response between 70~cm$^{-1}$ and 1000 cm$^{-1}$ (see insets of figure \ref{spectres}). Starting from T$\sim$300~K, the integrated response first increases until a temperature $T^{'}$ where it starts decreasing until we reach the superconducting transition. The characteristic temperature $T^{'}$ extracted from this analysis is at or above 200~K for the four samples. It does not display any obvious trend with doping although it is slightly higher in the optimal doped sample (Hg95K) where $T^{'}\sim$240~K. The characteristic frequency and temperature scales of the depletion are summarized in figure \ref{evolution} where they are plotted as a function of doping. In addition to $T^{'}$, we also show the evolution of the characteric energies of the depletion $\omega_1$ and $\omega_2$. $\omega_1$ is defined as the onset frequency of the depletion while $\omega_2$ is defined as the frequency where the depletion is maximum (see inset of figure \ref{evolution}). In the four samples studied both $\omega_1$ ans $\omega_2$ are, within experimental accuracy, temperature independent. More strikingly, they appear to be virtually doping independent: around 700~($\pm$50)~cm$^{-1}$ and 300~($\pm$50)~cm$^{-1}$ for $\omega_1$ and $\omega_2$ respectively. Our data thus strongly suggest the existence of an ubiquitous energy scale in the normal state of Hg-1201 over a wide range of doping. We note that a similar depletion has been reported in underdoped Y-123 and Bi-2212 at similar characteristic frequencies and temperatures, thus suggesting an universal behavior among hole-doped cuprates \cite{Nemetschek}.

\begin{figure}[]
\centering
\epsfig{figure=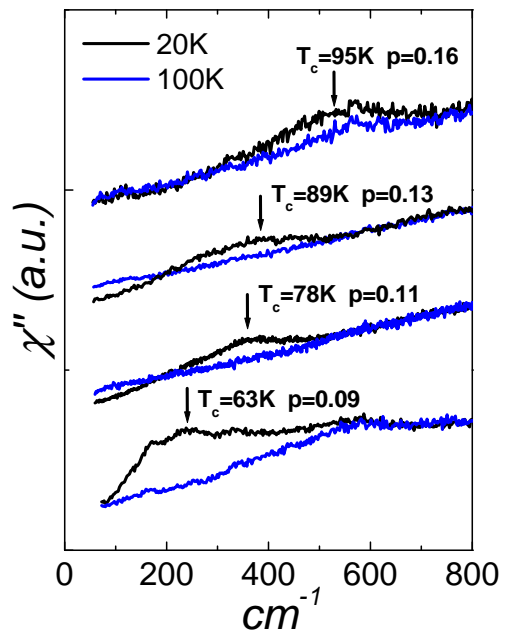, width=0.75\linewidth, clip=}
\caption{B$_{2g}$ responses both above and below $T_c$ as a function of doping in HgBa$_2$CuO$_{4+\delta}$. The pair-breaking peaks are marked by arrows.}
\label{supra}
\end{figure}

Before adressing the origin of the observed behavior, it is interesting to compare the doping evolution of the extracted frequencies with that of the superconducting gap. The superconducting responses in the same symmetry are displayed in figure \ref{supra}. Contrary to the B$_{1g}$ symmetry where the superconducting response is strongly suppressed in underdoped samples \cite{Gallais2005}, a B$_{2g}$ superconducting response is observed at all doping as evidenced by the presence of a pair-breaking peak below $T_c$. The position of the peak is strongly doping dependant going from 240~cm$^{-1}$ for Hg63K up to 520 cm$^{-1}$ for Hg95K implying that the amplitude of the $d$-wave gap near the nodes decreases towards lower doping. While this is a common trend observed in the B$_{2g}$ Raman response of most hole-doped cuprates (see for example ref. \cite{venturini}), it contradicts the analysis of thermal conductivity where the opposite trend is found \cite{sutherland}. This very important issue remains a puzzle because experiments which probe the nodal qps are expected to be less affected by the pseudogap and should therefore yield reliable estimates of the superconducting gap.

\begin{figure}
\centering
\epsfig{figure=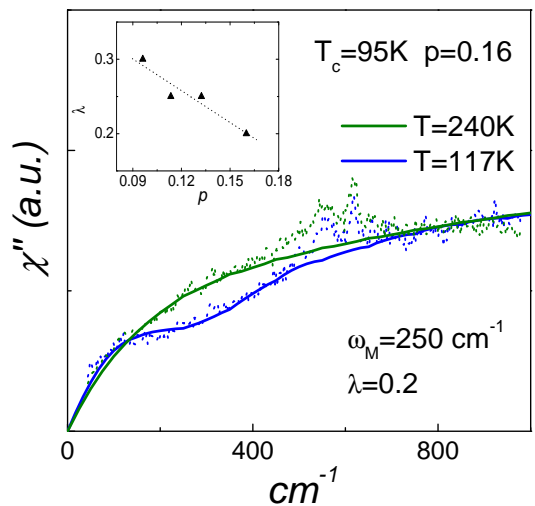, width=0.8\linewidth, clip=}
\caption{Theoretical fit of the Raman response using the electron-boson coupling model (see text). The fit has been made on the response of the optimally doped sample Hg95K at $T$=117~K. We also show a simple MFL fit of the response at higher temperature ($T$=240~K) where the effect of electron-boson coupling are expected to be smeared out. The depletion in the reponse around $\omega_M$ for the spectrum at $T$=117~K is well reproduced by our model. In the inset, the doping dependence of the coupling parameter $\lambda$ extracted from similar fit on the other samples.} 
\label{simul}
\end{figure}

While this issue remains to be addressed, the key point here for us is that the normal state characteristic frequencies and the superconducting gap have very different doping dependence, ruling out any possible precursor effect to explain the normal state depletion. It is however tempting to associate the depletion to the opening of a anisotropic pseudogap in the charge response. While previous data have been interpreted along these lines \cite{Gallais2005,Nemetschek}, the present study makes this interpretation less tenable. In particular the doping dependences of $T^{'}$, $\omega_1$ and $\omega_2$ contrast with that of the pseudogap whose temperature and energy increase strongly when doping is decreased \cite{ARPES-RMP}. 
Rather, we adopt the alternate point of view that the depletion is the result of the strong coupling between the qps and a collective bosonic mode. To make a case for our interpretation, we consider a model of ERS by qps coupled to a collective bosonic mode. The Raman response $\chi^{"}$ can be written in term of the spectral fonction $A(k,\omega)$:
\begin{multline}
\chi^{''}_{\gamma\gamma}(\textbf{q}=0,\Omega)=\frac{2}{N}\sum_{\textbf{k}}\gamma^2(\textbf{k})\int\frac{d\omega}{\pi}
[f(\omega)-f(\omega+\Omega)] \\
A(\textbf{k},\omega)A(\textbf{k},\omega+\Omega)
\label{reponse}
\end{multline}
where $\gamma$ is the Raman vertex, which in B$_{2g}$ symmetry is proportional to sin$(k_xa)$sin$(k_ya)$, $a$ is the lattice constant, $f$ is the Fermi-Dirac function and $A(k,\omega)$ can be written in term of the self-energy $\Sigma$:
\begin{equation}
\label{spectral}
A(\textbf{k},\omega)=-\frac{1}{\pi}\times\frac{\Sigma^{''}(\textbf{k},\omega)}{(\omega-\epsilon_{\textbf{k}}-\Sigma^{'}(\textbf{k},\omega))^2+(\Sigma^{"}(\textbf{k},\omega))^2}
\end{equation}
The self-energy $\Sigma$ for a coupled electron-boson system has been worked out by Engelsberg and Schrieffer for the case of a Debye phonon spectrum \cite{Engelsberg}. It is a very general result for the case of a coupling between electrons and collective bosonic excitations of characteristic frequency $\omega_M$. In this simple case, the real part of the electronic self-energy (or mass renormalization) has a peak at the frequency $\omega_M$ while its imaginary part (or relaxation rate) decays rapidly when $\omega<\omega_M$ and is constant for $\omega>\omega_M$. The real and imaginary parts of the self energy can be written as (at $T$=0~K) \cite{Engelsberg,Verga}:
\begin{multline}
\label{selfre}
\Sigma^{'}(\omega)=-\frac{\lambda\hbar\omega_M}{3}\times\biggl[\biggl(\frac{\omega}{\omega_M}\biggr)^3\ln\biggl|\frac{(\omega^{2}_{M}-\omega^2)}{\omega^2}\biggr| \\
+\ln\biggl|\frac{\omega_M+\omega}{\omega_M-\omega}\biggr|+\frac{\omega}{\omega_M}\biggr]
\end{multline}
\begin{eqnarray}
\label{selfim}
|\Sigma^{''}(\textbf{k},\omega)|=\hbar\lambda\pi\frac{|\omega|^3}{3\omega_{M}^{2}}, & \quad |\omega|<\omega_M \nonumber\\
|\Sigma^{''}(\textbf{k},\omega)|=\hbar\lambda\pi\frac{\omega_M}{3}, & \quad |\omega|>\omega_M 
\end{eqnarray}
where $\lambda$ is a coupling constant.   
The above expression has been used for the analysis of ARPES data showing strong electron-phonon coupling at the surface of Be \cite{Lashell}. The $T$=0~K approximation is justified as long as the temperature is considerably smaller than the mode energy itself. In cuprates, a more realistic model must include electron-electron interaction which results in an additional frequency dependence of the scattering rate. We account for this by adding to the above self-energy a marginal Fermi liquid-like (MFL) component  $\Sigma_{MFL}(\textbf{k},\omega)\sim \omega\ln\frac{x}{\omega_c}-i\frac{\pi}{2}x$ with $x=max(|\omega|,T)$ and $\omega_c$ a UV cutoff \cite{varma}. While necessary to account for the non-vanishing ERS continuum at high frequency, the inclusion of this term is, however, not crucial for our analysis. The key point here is the absence of any energy scale except the temperature in the MFL phenomenology which leaves $\omega_M$ as the only relevant low-energy scale in the problem. The calculated response has been fitted to the experimental response of Hg95K at $T$=117~K by variyng $\omega_M$, $\lambda$ and using typical MFL parameters to fit the high frequency background \cite{Gallais2005}. The result is displayed in figure \ref{simul}. The good agreement with the experimental data enables us to extract the values of $\lambda$ and $\omega_M$. At high temperature (k$_BT\sim\hbar\omega_M$), the effect of the mode is smeared out and a MFL-like behavior is recovered as shown for the spectra at $T$=240~K in figure \ref{simul}. Similar fits were performed on the other samples and good agreements were found. The coupling parameters $\lambda$ extracted from the fits are plotted as a function of doping in the inset of figure \ref{simul}. They show an increase towards underdoping, similar to the trend found in the analysis of ARPES results \cite{shen-nagaosa}. We note, however, that our values of coupling parameter are expected to be underestimated due to the fact that we have ignored the effect of temperature on the self-energy. It should be noted that the value $\lambda\sim$1 has been deduced from density functional theory on Hg-1201 ($\Delta=0$), in fair agreement with our result condidering the simplicity of our model \cite{Ambrosch}. The mode frequency, on the other hand, shows hardly any doping dependence and is found to be close to the value $\omega_2$ determined experimentally (see figure \ref{evolution}). 
It gives a mode frequency of about 300$\pm$60~cm$^{-1}$ (38$\pm$ 8~meV) which is somewhat lower than the energies extracted from the normal state ARPES spectra. This frequency is in fact close to the spin resonance observed by Inelastic Neutron Scattering. However, the neutron spin resonance mode is mostly present in the superconducting state and is known to scale with $T_c$ \cite{bourges,Dai}. It is therefore an unlikely candidate to explain our data. In fact while a spin fluctuation continuum is still observed in the normal state and cannot be ruled out, we believe the very weak doping dependence of $\omega_M$ makes the phonon interpretation more natural. Recently, anomalies in the Cu-O bond stretching phonon dispersion were found in Hg-1201 \cite{neutron-Hg}. In particular a strong softening towards the zone boundary was observed around 50~meV and could provide a candidate for the energy scale observed in the ERS data. 

In summary, we have shown the existence of an energy scale in the normal state excitation spectrum of Hg-1201 over a wide doping range. We have provided a simple and natural picture of this observation as the consequence of coupling between the quasiparticles and a collective bosonic mode. The identification of this mode as a phonon seems the most convincing picture. Additional data on other cuprates may help resolve this issue.

We are grateful to T.P. Devereaux for useful discussions.

\end{document}